# Shear Instabilities in Granular Flows down on Inclined Plane

Hua-Shu Dou, Boo Cheong Khoo, Nhan Phan-Thien
Fluid Mechanics Division
Department of Mechanical Engineering
National University of Singapore
Singapore 119260
Email: tsldh@nus.edu.sg; huashudou@yahoo.com

**Abstract** Instabilities at interface of two stream granular flows have been reported in recent experiment [1] that breaking waves can form at the interface between two streams of identical grains flowing on an inclined plane downstream of a splitter plate. In this report, the theory of hydrodynamic instability is used to analyze the shear flow of granular materials. It is shown that the shear instability in two-stream granular flows actually comes from the competition between the energy gradients in transverse and streamwise directions as well as the interaction of two streams. We argue that the flow energy loss in the streamwise direction has a stabilizing effect, while the transverse component of the friction force formed by grain surface friction acts as the source of instabilities. An equation has been derived to characterize the transition between steady and wavy flows. Good qualitative agreement with the experimental data is obtained.

**Key Words:** Shear Instability; Granular flows; Incline; Gravity; Energy Gradient.

## 1. INTRODUCTION

Granular flows have displayed a variety of instabilities under various flow and geometric conditions which are responsible for the phenomena of mixing, segregation, jamming, earthquakes and faulting etc [1-8]. Recently, Goldfarb et al [1] reported a study on wave instability in granular flows down on an inclined chute (Fig.1a). They observed experimentally that breaking waves can form at the interface between two streams of identical grains flowing on an inclined plane downstream of a splitter plate. Changes in either the shear rate or the angle of incline cause such waves to appear abruptly. The instability is more intensive as the inclined angle of the plate to the horizontal is decreased. They suggested that the wave results from the competition contention between shear and extensional strains in the flowing granular bed, and proposed a dimensionless shear number to express the extent for transition between steady and wavy flows. The proposed mechanism and the criterion, however, could not be used to explain certain observed phenomena. For example, the sheared flow with glass beads does not generate the supposedly wave instability, while the competition between shear and elongation still exists. Flows with unsieved sands generate more extensive waves than the sieved sands, and but the competition of the shear and elongation may act equally with the sieved sands and unsieved sands. These unanswered issues provided the motivation for this study.

Dou (2002) [9] proposed an energy gradient theory with the aim to clarify the mechanism of transition from laminar to turbulence. It is thought that the gradient of total energy in the transverse direction of the main flow and the viscous friction in the streamwise direction dominate the instability phenomena and hence the flow transition for given disturbance. The energy gradient in the transverse direction has the potential to amplify a small velocity disturbance, while the viscous friction loss in the streamwise direction can resist and absorb this small disturbance energy. The transition to turbulence depends on the relative magnitude of the two roles of energy gradient amplifying and viscous friction damping to the initial disturbance. Based on such, a new dimensionless parameter, $K$ (the ratio of the energy gradient in the transverse direction to that in the streamwise direction), can be written as,

$$K = \frac{\partial E/\partial n}{\partial E/\partial s}. \tag{1}$$

Here, $E = p + \frac{1}{2}\rho V^2 + \rho g \xi$ is the total energy for incompressible flows with $\xi$ as the coordinate perpendicular to the ground, $n$ denotes the direction normal to the streamwise direction and $s$ denotes the streamwise direction. The parameter $K$ in Eq.(1) is a field variable. The occurrence of instability depends upon the magnitude of this dimensionless parameter $K$ and the critical condition is determined by the maximum value of $K$ in the flow, $K_{max}$. For a given flow geometry and fluid properties, when $K_{max}$ in the flow field is larger than a critical value, it is expected that instability would occur for certain initial disturbance [9]. The analysis show that the transition to turbulence is due to the energy gradient and the disturbance amplification, rather than a linear eigenvalue instability type [10-12]. Dou (2002) demonstrated that the criterion has obtained excellent agreement with the experimental data for parallel flows. The critical value of $K_{max}$ is about 370-380 for plane Poiseuille flow and pipe Poiseuille flow as well as plane Couette flow. Dou also suggested that this theory may be extended to the other areas such as flow instability in granular flows.

In this work, a simple and yet reasonably useful constitutive relation is proposed for slow flows of frictional granular materials under gravity. Then the energy gradient theory of hydrodynamic instability [9] is used to study and analyze sheared flows of granular material, and an equation to characterize the transition between steady and wavy flows is derived. The experiments observed phenomena in [1] are reasonably explained.

## 2. GOVERNING EQUATION AND CONSTITUTIVE MODEL UNDER GRAVITY

The conservation of momentum for an incompressible fluid can be expressed as,

$$\rho\left(\frac{\partial \mathbf{V}}{\partial t}+\mathbf{V}\cdot\nabla\mathbf{V}\right)=-\nabla p+\nabla\cdot\mathbf{T}+\mathbf{F}, \qquad (2)$$

where $t$ is the time, $\mathbf{V}$ the velocity vector, $\rho$ the density of the material, $p$ the hydrodynamic pressure, $\mathbf{T}$ the stress tensor, and $\mathbf{F}$ is the gravity force with gravitational acceleration $\mathbf{g}$.

Models of fluid mechanics like have been used in granular materials [13-21]. In some models, the stress among the particle of grains is considered to be composed of collision and friction mechanisms [2,14]. We reckoned that these two roles produce the apparent "viscosity" of the flow and the contact stress. The stress tensor for the flow of granular material is split into two parts, with analogy to Oldroyd-B model [22], $\mathbf{T}=\mathbf{T}_1+\mathbf{T}_2$; $\mathbf{T}_1=2\mu\mathbf{D}$ is the Newtonian stress tensor with $\mu$ as the dynamic viscosity of the fluid and $\mathbf{D}$ is the tensor of rate-of-strain, and $\mathbf{T}_2$ is the surface frictional stress tensor of grains. We consider that $\mu$ is the apparent viscosity of the fluid cell and comes primarily from the collision and deformation of the materials, and $\mathbf{T}_2$ comes mainly from the surface friction among particles. For frictionless material such as smoothed glass balls, $\mathbf{T}_2=0$ and $\mathbf{T}=\mathbf{T}_1$.

Introducing above relations into Eq.(1), we obtain,

$$\rho\left(\frac{\partial \mathbf{V}}{\partial t}+\mathbf{V}\cdot\nabla\mathbf{V}\right)=-\nabla p+\nabla\cdot(2\mu\mathbf{D})+\nabla\cdot\mathbf{T}_2+\rho\mathbf{g}. \qquad (3)$$

We assume that the flow of the material is driven by gravity alone. Thus, we consider that $\nabla \cdot \mathbf{T}_2$ is generated by the gravity force only. This should be reasonable especially so for the slower moving flows. Therefore, we have,

$$0 = \nabla \cdot \mathbf{T}_2 + \lambda \eta \rho \mathbf{g}, \qquad (4)$$

where $\lambda \eta < 1$, $\lambda > 0$, and $\eta > 0$. Here, $\lambda = \lambda(x, y)$ is a function of the coordinates related to the kinematics, and $\eta$ is the surface friction factor of the material.

For the velocity vector, there is the identity,

$$\mathbf{V} \cdot \nabla \mathbf{V} = \frac{1}{2} \nabla (\mathbf{V} \cdot \mathbf{V}) - \mathbf{V} \times \nabla \times \mathbf{V}. \qquad (5)$$

Substituting the Eq.(4) and (5) into Eq. (3), the following equation can be obtained,

$$\rho \frac{\partial \mathbf{V}}{\partial t} + \nabla \left( p + \frac{1}{2} \rho V^2 + \rho g \xi \right) = \mu \nabla^2 \mathbf{V} - \lambda \eta \rho \mathbf{g} + \rho \left( \mathbf{V} \times \nabla \times \mathbf{V} \right), \qquad (6)$$

where $\xi$ is the coordinate perpendicular to the ground. If we take the $(x, y)$ coordinates as in Fig.1b, we obtain for two-dimensional flow,

$$\rho \frac{\partial u}{\partial t} + \frac{\partial}{\partial x} \left( p + \frac{1}{2} \rho V^2 - \rho g \sin \alpha \, x \right) = \mu \left( \frac{\partial^2 u}{\partial x^2} + \frac{\partial^2 u}{\partial y^2} \right) - \lambda \eta \rho g \sin \alpha + \rho v \left( \frac{\partial v}{\partial x} - \frac{\partial u}{\partial y} \right), (7)$$

$$\rho \frac{\partial v}{\partial t} + \frac{\partial}{\partial y} \left( p + \frac{1}{2} \rho V^2 + \rho g \cos \alpha \, y \right) = \mu \left( \frac{\partial^2 v}{\partial x^2} + \frac{\partial^2 v}{\partial y^2} \right) + \lambda \eta \rho g \cos \alpha - \rho u \left( \frac{\partial v}{\partial x} - \frac{\partial u}{\partial y} \right). (8)$$

### 3. FLOW INSTABILITY

Next, we apply the theory of flow instability developed for turbulence transition [9] to the granular flows. We utilize the governing equations (7, 8) for the energy gradient to calculate the value of $K$ in the flow field. According to [9], the position with a maximum of $K$ in the flow should be the most "unstable" location if the instability appears.

For steady flow of the granular materials considered here (Fig. 1), and substituting Eq.(7) and (8) into Eq.(1) we have,

$$K = \frac{\frac{\partial}{\partial y}\left(p + \frac{1}{2}\rho V^2 + \rho g y \cos\alpha\right)}{\frac{\partial}{\partial x}\left(p + \frac{1}{2}\rho V^2 - \rho g x \sin\alpha\right)} = \frac{\lambda \eta \rho g \cos\alpha + \rho u \frac{\partial u}{\partial y}}{\mu\left(\frac{\partial^2 u}{\partial y^2}\right) - \lambda \eta \rho g \sin\alpha}. \qquad (9)$$

At low velocity, the fluid inertia is negligible and we obtain,

$$K = \frac{\lambda \eta g \cos\alpha}{\nu\left(\frac{\partial^2 u}{\partial y^2}\right) - \lambda \eta g \sin\alpha}, \qquad (10)$$

where $\nu = \mu/\rho$ is the kinetic viscosity. The above equation is simplified to

$$K = \frac{\lambda \eta g \cos\alpha}{(a - \lambda\eta) g \sin\alpha} = \frac{\lambda \eta}{(a - \lambda\eta)\tan\alpha}, \qquad (11)$$

where $a = \nu\left(\frac{\partial^2 u}{\partial y^2}\right)/g \sin\alpha$. The parameter $a$ is negative due to flow energy loss. For steady flow, $a = a(y)$ and $\lambda = \lambda(x, y) = \lambda(y)$, and hence $K$ is a function of $y$ only along the transversal direction. $K$ attains its maximum value at the position where $a$ is maximized. It should be mentioned that Eq.(11) is *exact* for inertialess flows regardless of how the frictional stress tensor $\mathbf{T}_2$ is constructed.

For two stream flows side by side shown as in Fig.1, we assume that the flow is two-dimensional in x-y plane for each flow and we have within each flow,

$$K = \frac{\lambda_1 \eta}{(a_1 - \lambda_1\eta)\tan\alpha} \text{ and } K = \frac{\lambda_2 \eta}{(a_2 - \lambda_2\eta)\tan\alpha}. \qquad (12)$$

Here the subscript 1 and 2 denote the two streams respectively (Fig.2). The velocity profiles of two streams are sketched in Fig.2b. At the interface, the two flows interact each other. At this interface, we obtain the maximum of $K$ of the flow field at some position y,

$$K_{max} = \frac{(|\lambda_1 \eta|, |\lambda_2 \eta|)_{max}}{(|a_1 - \lambda_1 \eta|, |a_2 - \lambda_2 \eta|)_{min} \tan \alpha}. \qquad (13)$$

As shown in Fig.2a, a band with high value of $K$ will be formed at the interface.

We made the following observations:

**(1) Interface instability**: For two-stream sheared flows, when the velocity difference between the two streams is small and the inclined angle is large, $K_{max}$ is small. Thus, the flow is stable and no wavy instability appears. When the velocity difference between the two streams becomes large and the inclined angle is reduced, $K_{max}$ increases. When $K_{max}$ reaches beyond a critical value, the flow instability may appear at interface following Eq.(13). This instability occurrence displays a three-dimensional behavior of the interaction, i.e, *the behaviour in x-y plane at interface results in the wavy instability in the third direction*. This is possibly the reason why the wavy interface instability is formed with interfacial shear of two streams. This phenomenon is different from the Kelvin-Helmholtz instability in free shear flows of Newtonian fluids. In Kelvin-Helmholtz instability, the instability is due to the formation of inflexion profile and the high Reynolds number. The instability can be considered as that it is induced by increasing the kinetic energy gradient in transverse direction. The flow can be only two-dimensional, and there is formation of vortices along the interface. In experiment [1], only wavy instability and wave breaking appear, and no vortex is found. The reason is explained as below. Although there is shear between the two streams, the kinetic energy gradient in the direction normal to x-y plane is small owing to the low Reynolds number. Therefore, vortices like in free shear flows could not be formed.

The effect of velocity profile on stresses has been studied in [23] and it was shown that the amplitude of avalanche depends on the velocity profile. Hartley and Behringer [24] also demonstrated that there is a logarithmic velocity profile in shear flows of granular matters, rather than the stresses being independent of the shear rate as in some studies (see [24]). This is consistent with the assumption in our proposed model that the stress is related to the shear rate.

**(2) Effect of inclined angle:** Equation (13) indicates that $K_{max}$ is inversely proportional to $\tan\alpha$. When $\alpha$ is reduced, $K_{max}$ increases. Thus, a smaller $\alpha$ leads to the flow being more unstable. In other words, making the inclined angle to the horizontal leads to a more intense instability, in agreement with experiments shown on wavy breaking when the angle is reduced [1]. We also noted that $K_{max}$ in Eq.(13) becomes singular when the inclined angle is reduced to zero. This should not happen because there is a critical value of $\alpha$ for a given material thickness below which the flow cannot be sustained [20, 25-26]. In [1], the critical value obtained is 19.6 degree.

Fig. 3 shows the possible comparison of theory with the experiments from [1]. When the surface frictional factor of grains is large, the effect of the parameter $a_1$ and $a_2$ on the value of $K$ is small. In this case, the value of $K$ mainly depends on the distribution of frictional force ($\lambda(x,y)\eta$). If we further assume that the value of $(|\lambda_1\eta|, |\lambda_2\eta|)_{max} / (|\lambda_1\eta|, |\lambda_2\eta|)_{min}$ is not strongly dependent on $\alpha$, we have $K_{max} \propto (\tan\alpha)^{-1}$ roughly from Eq.(13). This relation is provided in Fig. 3, together with the experimental data in [1]. Although it is not exactly that the wave amplitude is proportional to $K_{max}$, the trend is observed here.

**(3) Smoothed grains**: For smoothed grains, $\eta = 0$ and $K_{max} = 0$ in Eq.(13). Therefore, the interface instability will never occur. Glass balls are nearly frictionless and $\eta$ is very small (thus $K_{max}$ small), and instability should not occur with two-stream glass ball flows.

**(4) Unsieved sand versus sieved sand**: Unsieved sand contributes to the non-uniformity of grains, and leads to a large value of the apparent surface friction factor of grains. Thus, this will result in an increased $K_{max}$ according to Eq.(13). The two stream flows will be more unstable with unsieved sand.

**(5) Roughness of the bottom wall**: Experiments showed that the bottom wall roughness makes the flow more unstable [1]. Goldfarb et al [1] thought that it might be due to the reduced elongation of the flow. From the present theory, this phenomenon may be attributed to the following reasons. The roughness of the bottom wall results in disturbance to the flow and increases the friction of the media. Thus, the flow will be more unstable from Eq.(13).

It is clear that the reason and mechanism for the interfacial wavy instability of sheared flows can be broadly traced to the grain surface friction and the inclined angle as well as the velocity profile of the sheared flow. The grain surface friction generates normal stresses which lead to energy drop across the transversal direction. This energy gradient (drop) acts on the flow of materials transversely. These roles constructed the distribution of the parameter $K$ in x-y plane for each stream. The interaction of two-streams makes the value of $K$ large at interface. When the value of $K_{max}$ is larger than some critical value for a given flow condition, wavy instability may occur. Owing to the transversal energy gradient is proportional to the frictional factor, the grains with larger frictional factor will produce wavy instability easily. The discovery of the physics of shear instability will help the understanding of the flow of granular materials and to improve the constitutive relation development. These results have significant interests in the design of industrial processing and the prediction of catastrophic events.

## 4. CONCLUSION

An equation for the constitutive model ($\nabla \cdot \mathbf{T} = \nabla \cdot (2\mu \mathbf{D}) - \lambda(x,y)\eta\rho\mathbf{g}$) in Eq.(3) and (4) of granular flows under gravity is proposed with an analogy to the Oldroyd-B model for the frictional granular materials. This model gives the equilibrium relation for both smoothed and frictional materials. The instabilities at interface of two-stream granular flows can be described using the energy gradient theory which was proposed for laminar-turbulence transition. An equation has been derived for characterizing the extent of instability and the transition between steady and wavy flows. Good qualitative agreement with the experimental data is obtained. The argument of waves resulting from a competition between shear and extensional strains in the flowing granular bed as presented in [1] may not have provided a full picture; we show that the shear instability in granular flows comes from the competition between the energy gradients in the two directions (x,y) and the interaction between the two streams (leading to large $K$ value at interface). The wavy instability appearance displays a three-dimensional behaviour of the interaction, i.e., the behaviour in x-y plane at interface results in the wavy instability in the third direction.

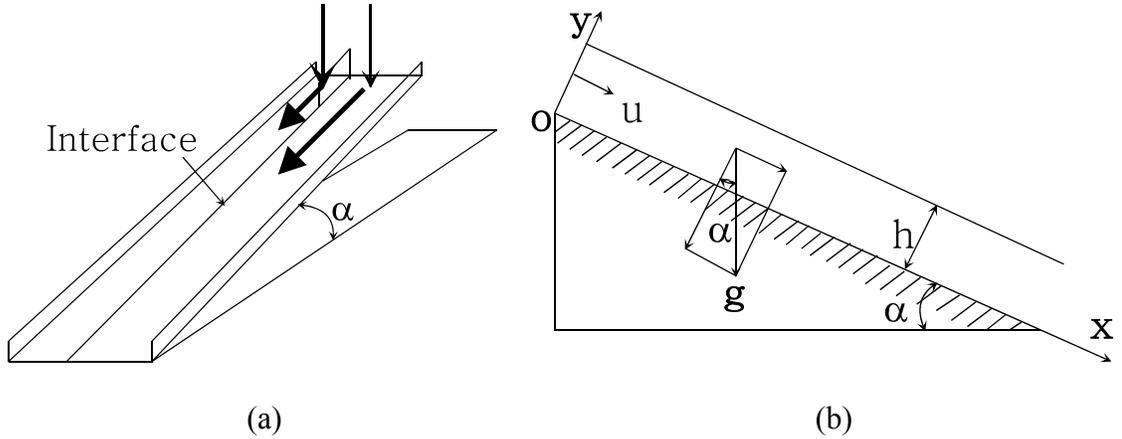

(a) (b)

Fig.1 Schematic of two-stream of flows down on inclined chute. (a) Flow geometry [1]; (b) Coordinates.

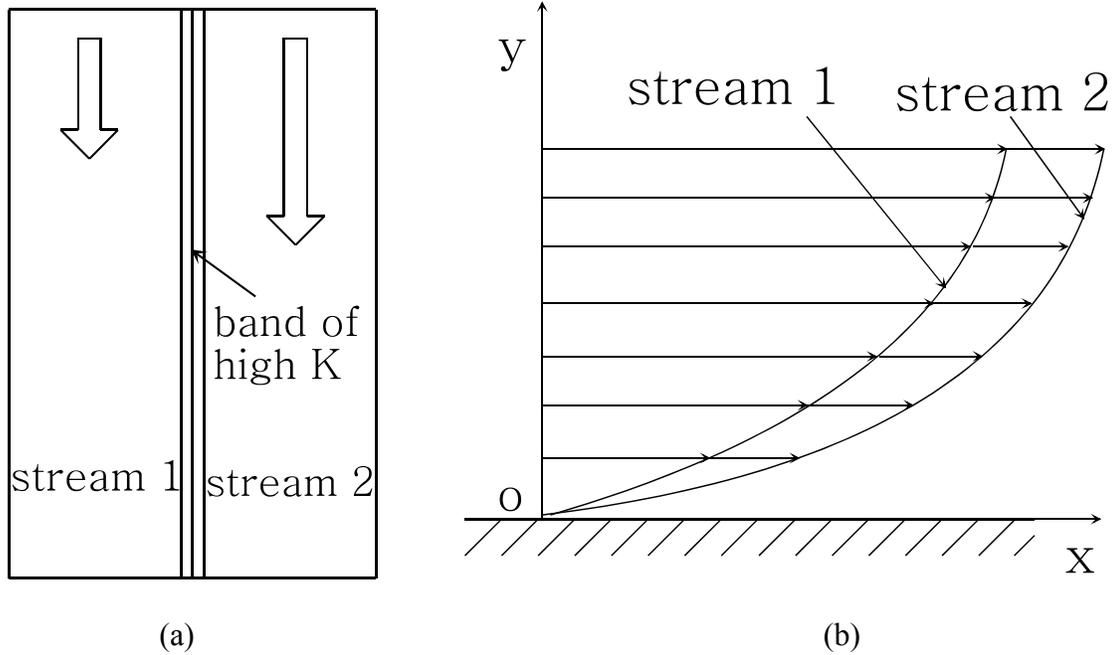

(a) (b)

Fig.2 (a) Band of high values of $K$ at interface of two-stream granular flows; (b) Velocity profiles of two streams.

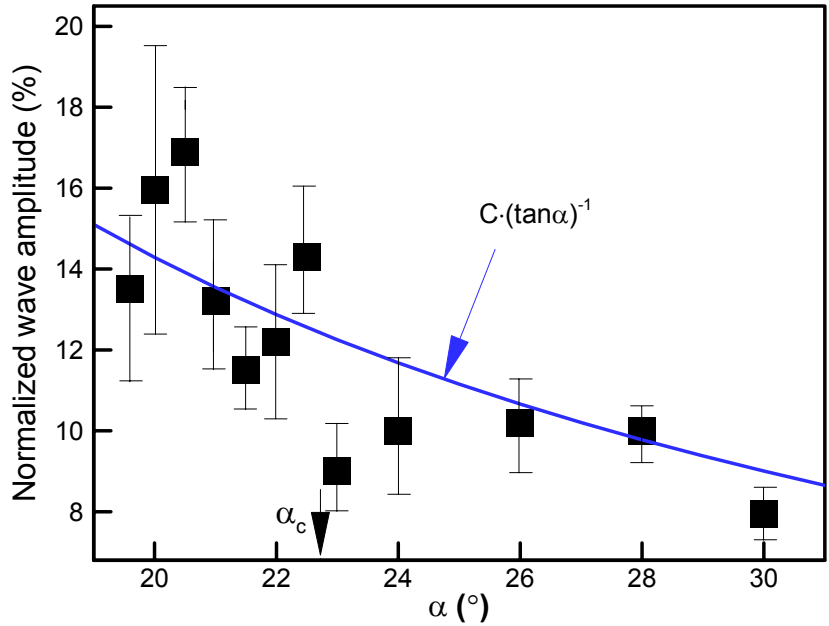

Fig.3 Comparison of the theory with the experimental data in [1].
Solid rectangular symbols: Experimental data for the wave amplitude versus the inclined angle $\alpha$. Solid line: $K_{max} \propto (\tan\alpha)^{-1} = C \cdot (\tan\alpha)^{-1}$ versus the inclined angle $\alpha$ (value also referring to the left ordinate), and the constant C used is 5.2. The value of $\alpha_c$ (about 23 degree) indicates the critical angle for the wave breaking in experiments. When value of $\alpha$ is less than 19.6 degree, the flow can not be sustained. Although it is not exactly that the wave amplitude is proportional to $K_{max}$, the same tendency of variations is clearly seen.

**References**
1. Goldfarb, DJ; Glasser, BJ; Shinbrot, T., , Nature, 415, 302-305 (2002).
2. Jaeger, H.M., & Nagel, S.R., Rev. Mod. Phys, 68, 1259-1273 (1996).
3. Goddard, JD, Annu Rev Fluid Mech., 35: 113-133 (2003).
4. Kadanoff, L. P. Built, Rev. Mod. Phys. 71, 435- 444 (1999).
5. Thorpe, S. A. , J. Fluid Mech. 39, 25-48 (1969).
6. Hansen,J.L., van Hecke, M., Haaning, A., Ellegaard, C., Andersen, K.H., Bohr, T., Sams, T., Nature 410, 324 (2001).
7. Wang, C.-H., Jackson, R. & Sundaresan, S. , J. Fluid Mech. 342, 179-197 (1997).
8. Shinbrot, T., Alexander A & Muzzio, F.J, Nature 397, 675-678 (1999).
9. H.-S. Dou, Energy Gradient Theory of Hydrodynamic Instability, Technical Report of National University of Technology, 2002. Also submitted to Proc. Roy. Soc. London A, 2003. Also will be presented at The 3rd Inter. Conf. on Nonlinear Science, June 30 – 2 July 2004, Singapore.
10. Schmid, P.J., & Henningson, D.S. Stability and transition in shear flows, (Springer, New York, 2001).
11. Trefethen, L.N., Trefethen, A.E., Reddy, S.C., Driscoll, T.A. , Science, 261, 578-684 (1993).
12. Grossmann, S., Reviews of modern physics, 72, 603-618 (2000).
13. Savage, S. B. & Hutter, K., J. Fluid Mech. 199, 177-215 (1989).
14. Johnson, P. C., Nott, P. & Jackson, R., J. Fluid Mech. 210, 501-535 (1990).
15. O.R. Walton, Mech. Mater. 16 , 239–247 (1993).
16. Schaeffer, D.G., J. Diff. Eqns. 66, 19-50 (1987).
17. Tardos, G.I., Powder Technol. 92, 61-74 (1997).
18. Greve, R, Koch, T., Hutter, K., Proc. R. Soc. Lond. A 445, 399–413 (1994).
19. Losert, W., Bocquet, L., Lubensky, T. C. & Gollub, J. P., Phys. Rev. Lett. 85, 1428-1431 (2000).
20. Louge, M. Y. & Keast, S. C., Phys. Fluids 13, 1213-1233 (2001).
21. Bocquet, L; Losert, W; Schalk, D; Lubensky, T. C. & Gollub, J. P., Phys Rev E, 65 (1): art. no.-011307 (2002)
22. Bird, RB, Armstrong, RC, & Hassager, O. Dynamics of Polymeric Liquids, Vol.1: Fluid Mechanics, 2nd ed.,Wiley, New York, 1987.
23. Aradian, A., Raphae, E., and Gennes, PG., , Phys Rev E, 60(2) 2009-2019 (1999)
24. Hartley R.R., & Behringer, R.P., Nature, 421, 928-931 (2003).
25. Silbert, LE; Ertas, D; Grest, GS; Halsey, TC, Levine, D, and Plimpton, SJ, Phys Rev E, 64 (5): art. no.-051302 Part 1, (2001).
26. Pouliquen, O., Scaling, Physics of Fluids, 11, 542-548 (1999).